\documentclass[aps,prl,preprint,groupedaddress]{revtex4}

\usepackage{graphics}
\usepackage{rotating}

\begin{document}

\title{Extraction of the $K\bar K$ isovector scattering length from 
 $pp\to dK^+\bar K^0$ data near
threshold}

\author{R. H. Lemmer$^{\ddag}$}

\affiliation{Institut f\"ur Kernphysik,
Forschungszentrum J{\"u}lich, D--52425 J{\"u}lich, Germany, and
Max--Planck--Institut f\"ur Kernphysik, Postfach 10 39 80,
D--69029, Heidelberg, Germany.}

\date{\today}

\begin{abstract}

The results of a recent experiment measuring the reaction
$pp\to dK^+\bar K^0$ near
threshold are interpreted in terms
of a spectator model that encapsulates the main features of
the observed $K^+\bar K^0$ invariant mass distribution.
A  $\chi^2$ fit to this data leads to an
imaginary part of the  isovector scattering length in the
$K\bar K$ channel of
$Im(a_1)=-(0.63\pm 0.24)$ fm. 
 We then use  the
Flatt\'e representation of the
scattering amplitude to infer a  value  $Re(a_1)= -(
0.02\pm 0.02)$ fm for the real part under the assumption that scaling
(Baru {\it et al.}\cite{Baru05}) is approximately satisfied.
 We show further that it is not possible to exclude the effects of
$\pi^+\eta$ to $K^+\bar K^0$ channel coupling within the context of our model.\\


\vspace{10cm}
Corresponding author: R H Lemmer

Electronic address: rh$_-$lemmer@mweb.co.za

\end{abstract}
\maketitle

 \subsection{1. Introduction}

  The reaction $pp\to d K^+\bar K^0$ has been
 measured at an excess energy of $Q$ = $46$ MeV with the spectrometer ANKE at the cooler synchrotron
 COSY-J\"ulich\cite{VK03}. This experiment has achieved a resolution of $1-8$ MeV
 for the invariant mass--spectrum of the $K\bar K$ system which
 considerably improves the data base. Therefore it is timely to investigate what
 information regarding the $K\bar K$ scattering length can be extracted
 from this
 data.

 A model-independent analysis of the angular distributions shows a dominance
 of an $s$  wave
 between the two kaons accompanied by a $p$  wave deuteron with respect to the
 mesons. This has been interpreted as evidence
 for a dominant production of the $a_0(980)$ near threshold
 \cite{VK03,HANHART04}. Detailed calculations of the reaction
 $pp\to d K^+\bar K^0$ performed
 by Grishina {\it et al.}\cite{GRISHINA04} confirm this conclusion. In particular
 the non--resonant $ K^+\bar K^0$
 production is shown to be small below excess energies
 of $Q$ = $100$ MeV. Fadeev calculations for the $K^-d$ reaction 
 have extracted a scattering length $a_{K^-d}=
 (1.80-i1.55)$ fm (using the Bethe sign convention here)\cite{Bah03}. A summary
 of $K^-d$ scattering lengths obtained by different authors can be found in
 Ref.\cite{SIBIRTSEV04} with $Re(a_{K^-d})$ ranging from $0.01$ to $1.92$ fm.

 Data obtained more recently at COSY 
  have been analysed by Sibirtsev {\it et al.}
 \cite{SIBIRTSEV04}.
 These authors have extracted limits for the
 $\bar K d$ scattering length from the data of Ref.~\cite{VK03} and found the
 magnitudes  of both the real and imaginary parts of this quantity to be less
 than $1.3$ fm.  A chiral approach to the reaction
 $pp\to d K^+\bar K^0$ 
  has been formulated by Oset, Oller and
 Meissner \cite{OSET01}
 that suggests a potentially large sensitivity of the data to the
 $\bar K d$ scattering
 length. A  study \cite{FELIX05} based on this formalism finds an  optimal fit
 to the COSY data for a purely imaginary $\bar Kd$ scattering length
 of $a_{\bar Kd}$ = $(0.0+i1.0)$ fm.

 The finding of a small real part of the scattering length $a_{\bar K d}$
 in Refs.\cite{SIBIRTSEV04,FELIX05} encourages us to
  explore the working assumption
   that the deuteron's role is simply that of 
 a spectator, and that the
 $K^+\bar K^0$ production in  $pp\to dK^+\bar K^0$ is dominated by 
 final state interactions in the meson--meson channels.
 Since the reaction $pp\to d\pi^+\eta$ has a lower threshold than 
 $pp\to dK^+\bar K^0$, both $\pi^+\eta$ production and $K^+\bar K^0$
 production have to be treated simultaneously. We therefore consider a coupled
 channel description of the two--meson final states $\pi^+\eta$ and
 $K^+\bar K^0$. In the vicinity of the $K^+\bar K^0$ threshold, one expects the
 $\pi^+\eta$ production to dominate. This allows one to make a simple estimate of the
 two kaon production via the chain $\pi^+\eta\to a^+_0\to K^+\bar K^0$.
 In order to include the other reaction chain
 $K^+\bar K^0\to a^+_0\to K^+\bar K^0$, one would need a model for the
 production operators. In view of this we assume that the first reaction chain is
 dominant, but explore the influence of the second chain empirically.
 This approach cannot make any
 predictions  about  the three additional measured angular
  distributions or the
 $d\bar K^0$ invariant mass distribution
 reported in Ref.\cite{VK03}.

\subsection{2. Spectator model calculation}

  The scattering amplitudes for $K^+\bar K^0\to K^+\bar K^0$
  and $\pi^+\eta\to K^+\bar K^0$ are related by two--channel
  unitarity.
 Therefore at low relative momentum $k$ in the two kaon channel both amplitudes can be expressed
 in terms
 of the $K\bar K$ isovector complex scattering length $a_1=\alpha_s-i\beta_s$.
 Calling these two channels $1$ and $2$ respectively, one has \cite{COHEN80}
 \begin{eqnarray}
 &&T_{11}= \eta\; e^{2i\delta_1}-1\approx -2ika_1(1-ika_1)
 \nonumber
 \\
 &&e^{-i\delta_{12}}T_{12}=i\sqrt{1-\eta^2}\approx
 2i\sqrt{\beta_s k}(1-\beta_s k)
  \label{e:T}
 \end{eqnarray}
 where $\delta_{1,2}$ are the phase shifts in the corresponding
 channels, $\delta_{12}=\delta_1+\delta_2$, and $\eta$ is the inelasticity.

 We now assume that the $s$ wave  single differential 
cross section  $d\sigma^{(s)}/dP_0$ for the $pp\to dK^+\bar K$
three--body decay  can be written as the proportionality
\begin{eqnarray}
d\sigma^{(s)}/dP_0\sim\int_{t_{min}}^{t_{max}} dt
|AT_{12}+BT_{11}|^2\approx C_sk|(1-\beta_s k)
+\rho e^{i\phi}\sqrt{\beta_s k}\;|^2
\Delta t
\label{e:mixture}
\end{eqnarray}
where $\rho\;exp(i\phi)=iB/A(1+i\alpha_s/\beta_s)\;exp(-i\delta_{12})$
summarizes the admixture of channel $1$ via the ratio of the production operators $A$ and $B$ for channels $1$
and $2$,
$\Delta t$  is equal  to
 the difference in limits of the phase space integration,
\begin{eqnarray}
 &&\Delta t=t_{max}-t_{min}\sim\sqrt{(s-(m_D+P_0)^2)(s-(m_D-P_0)^2)},
\label{e:Deltat}
\end{eqnarray}
 $s=(p_A+p_B)^2$  is the invariant mass of the colliding protons, and
 $C_s$ a coefficient
 of proportionality. In principle both $C_s$ and  $\rho\;exp(i\phi)$  will depend on
 specific details and kinematics of the production mechanism in the two channels
 in question. 
 We assume these factors to be slowly--varying for simplicity.
For the three--body collision 
 the $K\bar K$ invariant mass is restricted to the range from threshold
$(P_0)_{min}=(M_{K^+}+M_{\bar K^0})$ to $(P_0)_{max}=(\sqrt s-m_D)$ so that
 $\Delta t$ vanishes at the upper limit; $m_D$ is the deuteron mass. At the COSY
 proton beam  energy of $T_p=2.65$ GeV ($p_p=3.46$ GeV/c) at which the
 experiment was performed, these limits are
 $(P_0)_{min}=0.9914$ GeV and $(P_0)_{max}=1.038$ GeV resp.
 This restricts the
 $K^+\bar K^0$ and $\pi^+\eta$ CM momenta,
\begin{eqnarray}
&&k=\frac{1}{2}P_0\Big[1-\frac{(M_{K^+}-M_{\bar K^0})^2}{P^2_0}\Big]^{1/2}
\Big[1-\frac{(M_{K^+}+M_{\bar K^0})^2}{P^2_0}\Big]^{1/2} 
\nonumber
\\
&&p=\frac{1}{2}P_0\Big[1-\frac{(M_\pi-M_\eta)^2}{P^2_0}\Big]^{1/2}
\Big[1-\frac{(M_\pi+M_\eta)^2}{P^2_0}\Big]^{1/2} 
\label{e:p}
\end{eqnarray}
  to lie in the very limited intervals $0<k<0.15$ GeV
 and $0.33<p<0.36$ GeV respectively.
The $P_0$ dependence of the cross section is thus completely determined by 
 the variation of
$k$ and $\Delta t$  with $P_0$ as fixed by Eqs.~(\ref{e:p}) and 
(\ref{e:Deltat}).
The scale constant $C_s[\mu b\;GeV^{-4}]$  
controls the absolute value of 
the $s$--wave cross section while the parameter $\beta_s=-Im(a_1)$ determines
its skewness in the case of no admixture, $\rho=0$.

The  expression equivalent to Eq.~(\ref{e:mixture}) for $p$ waves
to leading order in $k$ is
\begin{eqnarray}
d\sigma^{(p)}/dP_0\approx C_p k^3\Delta t
\label{e:sigfitp}
\end{eqnarray}
 which for $C_p=150.6\mu b\; GeV^{-6}$ reproduces the corresponding
 contribution determined in \cite{VK03}.

\begin{figure}
\rotatebox{-90}{\includegraphics{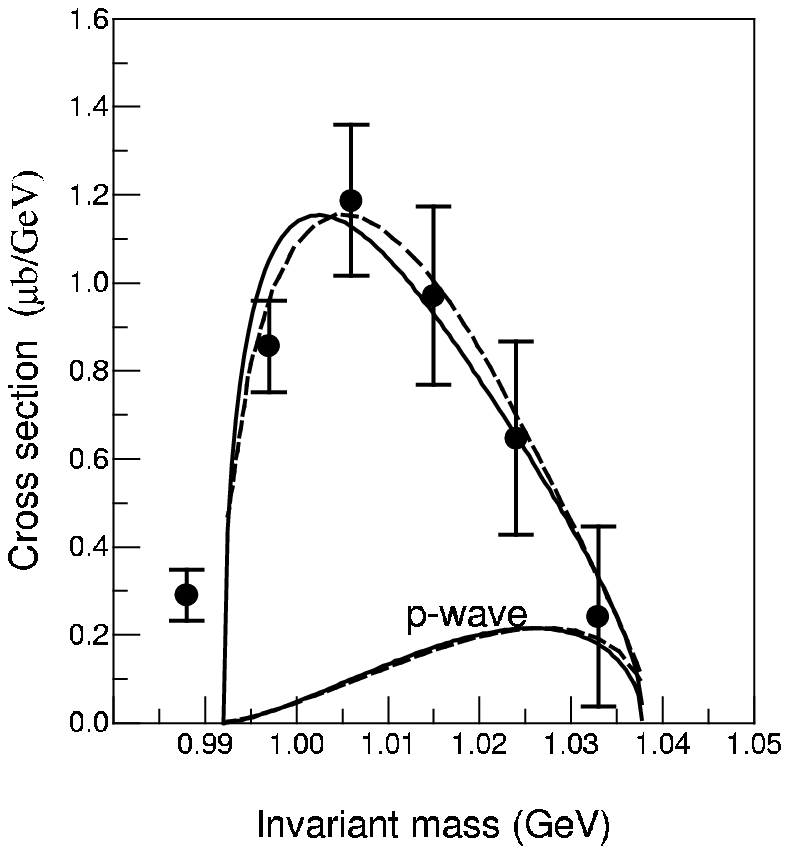}}
\caption{\label{f:fig1}
  The invariant mass 
 cross section for the production of $K^+\bar K^0$
 in $pp\to dK\bar K$.
 The solid line denotes a $\chi^2$ fit to the
 data\protect\cite{VK03} using the  spectator model without mixing
 (the $\chi^2$ fits in the presence of mixing are essentially identical
 for reasonable choices of the mixing parameters $[\rho,\phi]$).
 The broken curve shows the corresponding fit
  calculated in \protect\cite{VK03} using a six--parameter transition amplitude
 squared.  Shown separately are the $p$ wave contributions to the cross
 section obtained from
  Eq.~(\ref{e:sigfitp}) (solid curve),
 and the fit (broken curve) given in \protect\cite{VK03}.
  The error bars show the statistical uncertainties of the data.}

 \end{figure}

 The fits to the various data shown in Ref.\cite{VK03}
 employ  a six--parameter transition
matrix \cite{fits,VC02} fitted  
simultaneously to the two invariant mass distributions
$K^+\bar K^0$ and $d\bar K^0$,
plus three angular
distributions. In order to
compare the cross section given by the spectator model
with their result for the $K^+\bar K^0$ invariant mass distribution,
we have determined the  pair of
$s$ wave
constants $[C_s, \beta_s]$ by performing a $\chi^2$ fit
to the data points in
Fig.~\ref{f:fig1} that lie above threshold, using the
sum of Eqs.~(\ref{e:mixture}) and (\ref{e:sigfitp}) after expanding the former
up to order ${\cal O}(k)$,
initially for the case of no mixing,
$\rho=0$. In Sect.4 we discuss the effect of including channel admixtures
 as well as relaxing the assumptions on $C_s$.

  One finds $C_s=22.03\mu b\;  GeV^{-4}$, together with an
  imaginary part of the scattering length of
$\beta_s=Im(-a_1)=0.63$ fm at $\chi^2_{min}=0.65$.
 The  fit to experiment  for this  parameter set is shown
in
Fig. 1.
 By integrating out over the allowed range of $P_0$, one obtains a total cross 
section of
$\sigma(pp\to d K^+\bar K^0)=36$ nb, in agreement with
 the experimental cross section [$38\pm 2(stat)\pm 14(syst)]$ nb reported 
in \cite{VK03}.

\subsection{3. Flatt\'e  parametrization for scattering length}

The spectator model assumption Eq.~(\ref{e:mixture}) can only fix the
imaginary part of the scattering length for given [$\rho,\phi$]:
it does not allow one to extract the real
part directly. 
In order to make further progress
one therefore has to introduce specific model dependent assumptions regarding the structure of the
$K\bar K$ scattering amplitude in the isovector channel that describe the
formation and decay of the $a_0(980)$.  

There is little
experimental  information available on the $K\bar K$ scattering length.
However, within the last decade
the production of $\pi\eta$ pairs have been studied experimentally in
proton--antiproton annihilation, pion--proton reactions and in $\Phi$ decay,
and
analysed
using either Flatt\'e distributions\cite{NNA03,AAl02,ST99,MNA00}, or other
approaches\cite{DVB94,OBELIX03,AA98}.
 The  Flatt\'e  scattering amplitude\cite{flatte76}  
  near the  $K\bar K$ threshold takes on the form
\begin{eqnarray}
S_{K\bar K}\approx 1-\frac{i\bar g_{K\bar K} k}{-\epsilon_{a_0}+\frac{i}{2}
(\bar g_{K\bar K} k+\bar g_{\pi\eta}p)}
 \label{e:flat}
 \end{eqnarray}
 where $k$ and $p$ are the CM momenta defined in Eqs.~(\ref{e:p}), 
  while the $a_0(980)$ partial  decay widths $\Gamma_{\pi\eta}=\bar g_{\pi\eta}p$ and
 $\Gamma_{K\bar K}=\bar g_{K\bar K}k$ define the dimensionless coupling constants
 ($\bar g_{\pi\eta},\bar g_{K\bar K}$);
 $\epsilon_{a_0}=M_{a_0}-(M_{K^+}+M_{\bar K^0})$ is the (real or virtual)
 binding energy of the
 $a_0(980)$ relative to the $K\bar K$ threshold.  Eq.~(\ref{e:flat}) leads to
 the following result for the
   $K\bar K$ 
 isovector scattering length\cite{BM04}
\begin{eqnarray}
a_1=\frac{\frac{1}{2}\bar g_{K\bar K}}{-\epsilon_{a_0}+\frac{i}{2}
\Gamma_{\pi\eta}}=
\frac{R}{-\alpha+i}p^{-1}_{th}
\label{e:scatlength}
\end{eqnarray}
when rewritten in terms of the dimensionless ratios\cite{Baru05}
  $R=\bar g_{K\bar K}/\bar g_{\pi\eta}$
 and $\alpha=2\epsilon_{a_0}/\Gamma_{\pi\eta}$ instead of the individual
 coupling constants and energy parameters.
  Here $p_{th}=p[(P_0)_{min}]=0.33$ GeV = $1.65$ fm$^{-1}$ is the threshold momentum for $K\bar K$ production
  in the $\pi\eta$ channel.
    
 The $R$ and $\alpha$ are seen to be invariant
 under a scaling $\bar g_{K\bar K}\to \lambda \bar g_{K\bar K}$ etc. of the
  original Flatt\'e parameters. Table I shows a  compendium
   of mass, width
  and coupling constant values for $a_0(980)$ taken from the recent
  literature\cite{NNA03,AAl02,ST99,MNA00,DVB94,OBELIX03,AA98} and
  as partially summarised in \cite{BM04,Baru05}. We have also used
  Eq.~(\ref{e:scatlength}) to extend the calculations given in \cite{BM04}
  for the scattering lengths associated with these paramaters. The results
  are
  displayed in order of increasing $\alpha$ in Table I.

   From Eq.~(\ref{e:scatlength}) one sees that there is a
     linear relation between the real and imaginary parts of the (Flatt\'e)
     scattering lengths,
  \begin{eqnarray}
  Re(-a_1)=Im(-a_1)\alpha
   \label{e:linear}
  \end{eqnarray}
   If we now take $Im(-a_1)=0.63$ fm
  from the spectator model fit, this fixes the slope of the straight line in
  Eq.~(\ref{e:linear}). The result 
  can  then be compared  with  the values of $Re(-a_1)$ calculated from
  Eq.~(\ref{e:scatlength}) at the
   values of $\alpha$ and $R$ given in Table~I.
     This comparison is shown
  in Fig.~2. The spectator model result is seen to be consistent
  with  all of the data (except perhaps that of Ref. \cite{OBELIX03}) once the 
  uncertainties in $\alpha$ and hence $Re(-a_1)$ due to the experimental
  error in the
  $a_0(980)$ mass  are taken into consideration.

  It has also been pointed out\cite{Baru05} that 
   direct measurements of
  the $K\bar K$ elastic scattering cross section near threshold would serve to
  determine both $R$ and $\alpha$ independently, and thus the scattering length as given by
  Eq.~(\ref{e:scatlength}). Such data is not available yet. However,
  by assuming that scale invariance is valid,
  one can infer a value for $\alpha$ to be used in Eq.~(\ref{e:linear})
   by arguing as follows:
  the combinations
   $\alpha$ in Table I
  would have appeared as a common constant $\alpha_c$ had the data satisfied
  scaling  exactly. It is therefore reasonable to estimate this unknown
  constant by the 
   mean of the
  $\alpha$'s given in Table I, weighted by the inverse of the average error
  introduced into each $\alpha$ by the uncertainty in the $a_0(980)$ mass
  value used to
  calculate it. The resultant mean and its standard deviation is then
   $\alpha_c= 0.031\pm 0.054$. By also considering the changes
   in slope of the linear relation Eq.(\ref{e:linear}) due to the errors
   introduced
      into $Re(-a_1)$ by the mass value uncertainty one obtains the order
      of magnitude
   estimate  $\beta_s=(0.63\pm 0.24$) fm for the error in $\beta_s$.  
        Inserting these values for $\beta_s$ and $\alpha_c$ together with
   their estimated errors
   into Eq.~(\ref{e:linear}) gives 
         \begin{eqnarray}
      a_1\approx -[(0.02\pm 0.02)+i(0.63\pm 0.24)] {\rm fm}
      \label{e:scatfnl}
      \end{eqnarray}
  for the  isovector
  scattering length.

Note that this result relies on  two very different sources of information:
(i) a direct determination of $Im(a_1)$ based on the spectator model
fit of the $pp\to d K^+\bar K^0$ data for $\rho=0$, and (ii), an inferred estimate of
$Re(a_1)$ based on the assumption that scaling is approximately valid for
the experimental parameters extracted assuming a 
Flatt\'e distribution. 

\subsection{4. Effect of mixing with $K^+\bar K^0\to K^+\bar K^0$}
   
 We now repeat the $\chi^2$ fit of the data points in Fig.~1 in the
 presence of mixing.  Since $\rho>1$ can lead to unphysical values
 of $\beta_s<0$, we illustrate the effects of mixing for the two limiting
 cases
  labelled $[\rho,\phi]$ = $[1,\pi]$ and $[1,0]$
 in Fig.~2. The corresponding values of $\beta_s$ swing from $0.21$
 to $7.6$ fm respectively as illustrated by the gradients of these two straight
 lines.  However, by choosing an intermediate value of $\phi$ appropriately
 in the interval $0\leq \phi\leq \pi$ for given $\rho$,
 in this case $[\rho,\phi]=[1,110^0\pm 8^0]$, one again recovers 
  a fit ($\chi^2_{min}=0.82$) to the data in Fig.1 at the indicated value
  of $\beta_s$  given in Eq.~(\ref{e:scatfnl})
  that is essentially indistinguishable from the
  no--mixing case. Similar observations hold for
  any $0 < \rho < 1$ except that the swing in possible values of
  $\beta_s$ is smaller.

   Another source of uncertainty is associated with the lack of a detailed
   description of the production mechanism itself.
   As an illustration of this we compare
  with the specific model for the production mechanism suggested in Ref.\cite{VC02}
    that  codifies  selection rules explicitly.
  One finds that their scattering amplitude,
  which also considers
  resonance production, reduces  to   the first term
  of Eq.~(\ref{e:mixture}) at low momentum with
   $C_s\to (C^{'}_{s} p^2_D)$,
   where $p_D$ is the deuteron momentum.
    Making this replacement in Eq.~(\ref{e:mixture}),
    one obtains an equally
    good fit
    $(\chi^2_{min}=0.20)$ of the mass distribution
     in Fig.1  with  $\beta_s=0.63$ fm  by changing the mixing angle
    from $110^0$ to
    $59^0\pm 4^0$.
  
     At the level of the spectator model as
  envisaged in this note it is therefore impossible to decide whether or not
  channel mixing is important without introducing  model specific assumptions
  for the production operators $A$ and $B$.

   \subsection{Acknowledgments}
    We would like to thank the Ernest Oppenheimer Memorial Trust
    for research support in the form of  a Harry Oppenheimer Fellowship.
    Thanks are also due to
     the Institut f\"ur Kernphysik, Forschungszentrum, J\"ulich, and the
     Max--Planck--Institut f\"ur Kernphysik, Heidelberg for their
     kind hospitality. Discussions with M. B\"uscher and
     S. Krewald are gratefully acknowledged.
\newpage

\subsection{References}

\noindent
($^\ddag$) Permanent address: School of Physics,
 University of the Witwatersrand, Johannesburg,\\ Private Bag 3,
 WITS 2050, South Africa.

\newpage

 \begin{figure}
\rotatebox{-90}{\includegraphics{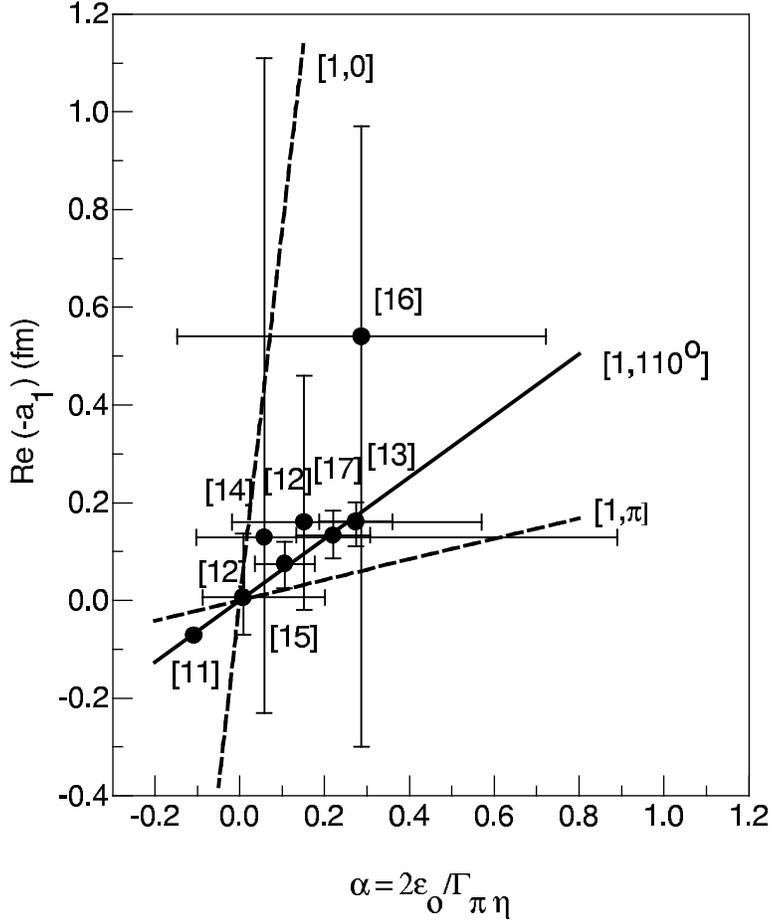}}
\caption{\label{f:fig2}
Plot of  $Re(-a_1)$ (filled circles) versus the dimensionless parameter
$\alpha=2\varepsilon_0/\Gamma_{\pi\eta}$. The plotted points
have been tagged by the corresponding  references in Table I. The horizontal and vertical
error bars show the errors introduced into $\alpha$, and hence $Re(-a_1)$,
by the uncertainties in the associated $a_0(980)$ mass values. 
The heavy straight
line
shows the calculated value of $Re(-a_1)$ using Eq.~(\ref{e:linear})
with $\beta_s=Im(-a_1)=0.63$ fm. The label $[1,110^0]$ gives the mixing
parameters that lead to the same value of $\beta_s$
upon including channel mixing as characterised by the parameters $[\rho,\phi]$
 defined in the text, while the remaining two broken straight lines labelled $[1,\pi]$ and $[1,0]$ show
the minimum and maximum values that  $\beta_s$ can assume for $\rho=1$.}

\end{figure}

\newpage

\begin{table}[ht]
\caption{\label{t:table1}
Summary of calculated values of the isovector scattering length
$a_1$   taken from Eq.~(\ref{e:scatlength}),
arranged in order of increasing  $\alpha=2\varepsilon_0/\Gamma_{\pi\eta}$,
for the various
mass, width and coupling constants for the $a_0(980)$ as
extracted\cite{Baru05} from the literature.}

\begin{ruledtabular}
\begin{tabular}{clrccccrr}
$Ref.$ & $M_{a_0}$ (MeV) & $\varepsilon_{a_0}$ (MeV) & $\Gamma_{\pi\eta}$ (MeV)
& $ \bar g_{\pi\eta}$ 
& $\bar g_{K\bar K}$ & $R$ & $\alpha$ & $a_1$ (fm)\\

\colrule

 \cite{AAl02} & $984.8^{+1.2}_{-1.2}$ & $-6.6$  & $121.5$ & $0.373$  & $0.412$
 & $1.10$
 & $-0.108$ & $+0.071-0.66i$\\

 \cite{NNA03} & $992^{+14}_{-7}$ & $0.6$ & $145.3$ & $ 0.446$ & $0.560$  &  $1.252$
 & $0.008$ & $-0.006-0.76i$\\

 \cite{MNA00} & $995^{+52}_{-10}$ & $3.6$  & $125$ & $0.384$ & $1.414$  & $3.685$
 & $0.058$ & $-0.13-2.23i$\\

   \cite{DVB94} & $999^{+5}_{-5}$ & $7.6$ & $143$ & $0.439$ & $0.516$  & $1.176$
  & $0.106$ & $-0.075-0.70i$\\

   \cite{NNA03} & $1003^{+32}_{-13}$ &  $11.6$ & $153$ & $0.470$ & $0.834$ & $1.776$
     & $ 0.152$ & $-0.16-1.05i$\\

      \cite{AA98} & $999^{+3}_{-3}$ & $7.6$ & $69$ & $0.212$ & $0.222$  & $1.048$
    & $ 0.220$ & $-0.13-0.61i$\\

   \cite{ST99} & $1001^{+3}_{-3}$ & $9.6$  & $70$ & $0.215$ & $0.224$  & $1.043$
   & $0.274$ & $-0.16-0.59i$\\                                       
   \cite{OBELIX03} & $998^{+10}_{-10}$ & $6.6$  & $46$ & $0.141$ & $0.476$  &
   $3.376$   & $0.287$ & $-0.54-1.89i$\\

 \end{tabular}
\end{ruledtabular}
\end{table}

\end{document}